\documentclass[sigconf]{acmart}
\usepackage{balance}

\AtBeginDocument{%
  }

\copyrightyear{2026}
\acmYear{2026}
\setcopyright{cc}
\setcctype{by}
\acmConference[SIGIR '26]{Proceedings of the 49th International ACM SIGIR Conference on Research and Development in Information Retrieval}{July 20--24, 2026}{Melbourne, VIC, Australia}
\acmBooktitle{Proceedings of the 49th International ACM SIGIR Conference on Research and Development in Information Retrieval (SIGIR '26), July 20--24, 2026, Melbourne, VIC, Australia}
\acmDOI{10.1145/3805712.3808504}
\acmISBN{979-8-4007-2599-9/2026/07}

\usepackage{multirow}
\usepackage{graphicx}
\usepackage{enumitem}

\begin{document}

\title{CASE: Cadence-Aware Set Encoding for Large-Scale Next Basket Repurchase Recommendation}

\author{Yanan Cao}
\authornote{Highlighted authors contributed equally to this research.}
\affiliation{%
  \institution{Walmart Global Tech}
  \city{Sunnyvale}
  \state{CA}
  \country{USA}
}
\email{yanan.cao@walmart.com}

\author{Ashish Ranjan}
\authornotemark[1]
\affiliation{%
  \institution{Walmart Global Tech}
  \city{Sunnyvale}
  \state{CA}
  \country{USA}
}
\email{ashish.ranjan0@walmart.com}

\author{Sinduja Subramaniam}
\affiliation{%
  \institution{Walmart Global Tech}
  \city{Sunnyvale}
  \state{CA}
  \country{USA}
}
\email{sinduja.subramaniam@walmart.com}

\author{Evren Korpeoglu}
\affiliation{%
  \institution{Walmart Global Tech}
  \city{Sunnyvale}
  \state{CA}
  \country{USA}
}
\email{ekorpeoglu@walmart.com}

\author{Kaushiki Nag}
\affiliation{%
  \institution{Walmart Global Tech}
  \city{Sunnyvale}
  \state{CA}
  \country{USA}
}
\email{kaushiki.nag@walmart.com}

\author{Kannan Achan}
\affiliation{%
  \institution{Walmart Global Tech}
  \city{Sunnyvale}
  \state{CA}
  \country{USA}
}
\email{kannan.achan@walmart.com}

\renewcommand{\shortauthors}{Yanan Cao et al.}

\begin{abstract}
Repurchase behavior is a primary signal in large-scale retail recommendation, particularly in categories with frequent replenishment: many items in a user’s next basket were previously purchased, and their timing follows stable, item-specific cadences. Yet most next basket repurchase recommendation models represent history as a sequence of discrete basket events indexed by visit order, which cannot explicitly model elapsed calendar time or update item rankings as days pass between purchases. We present \textbf{CASE} (\textbf{C}adence-\textbf{A}ware \textbf{S}et \textbf{E}ncoding) for next basket repurchase recommendation, which decouples item-level cadence learning from cross-item interaction, enabling explicit calendar-time modeling while remaining production-scalable. CASE represents each item’s purchase history as a calendar-time signal over a fixed horizon, applies shared multi-scale temporal convolutions to capture recurring rhythms, and uses induced set attention to model cross-item dependencies with sub-quadratic complexity, allowing efficient batch inference at scale. Across three public benchmarks and a proprietary dataset, CASE consistently improves precision, recall, and NDCG at multiple cutoffs compared to strong next basket recommendation baselines. In a production-scale evaluation with tens of millions of users and a large item catalog, CASE achieves up to 8.6\% relative precision lift and 9.9\% relative recall lift at top-5, showing that scalable cadence-aware modeling yields measurable gains in both benchmark and industrial settings.
\end{abstract}

%    <concept>
%        <concept_id>10002951.10003317.10003347.10003350</concept_id>
%        <concept_desc>Information systems~Recommender systems</concept_desc>
%        <concept_significance>500</concept_significance>
%        </concept>
%    <concept>
%        <concept_id>10010147.10010257</concept_id>
%        <concept_desc>Computing methodologies~Machine learning</concept_desc>
%        <concept_significance>500</concept_significance>
%        </concept>
%    <concept>
%        <concept_id>10002951.10003317.10003338</concept_id>
%        <concept_desc>Information systems~Retrieval models and ranking</concept_desc>
%        <concept_significance>300</concept_significance>
%        </concept>
%  </ccs2012>

\ccsdesc[500]{Information systems~Recommender systems}
\ccsdesc[500]{Computing methodologies~Machine learning}
\ccsdesc[300]{Information systems~Retrieval models and ranking}
\keywords{Next Basket Recommendation, Temporal Set Prediction, Sequential Modeling}


\maketitle

\section{Introduction}

\begin{figure*}[t]
    \centering    
    \includegraphics[width=1\linewidth]{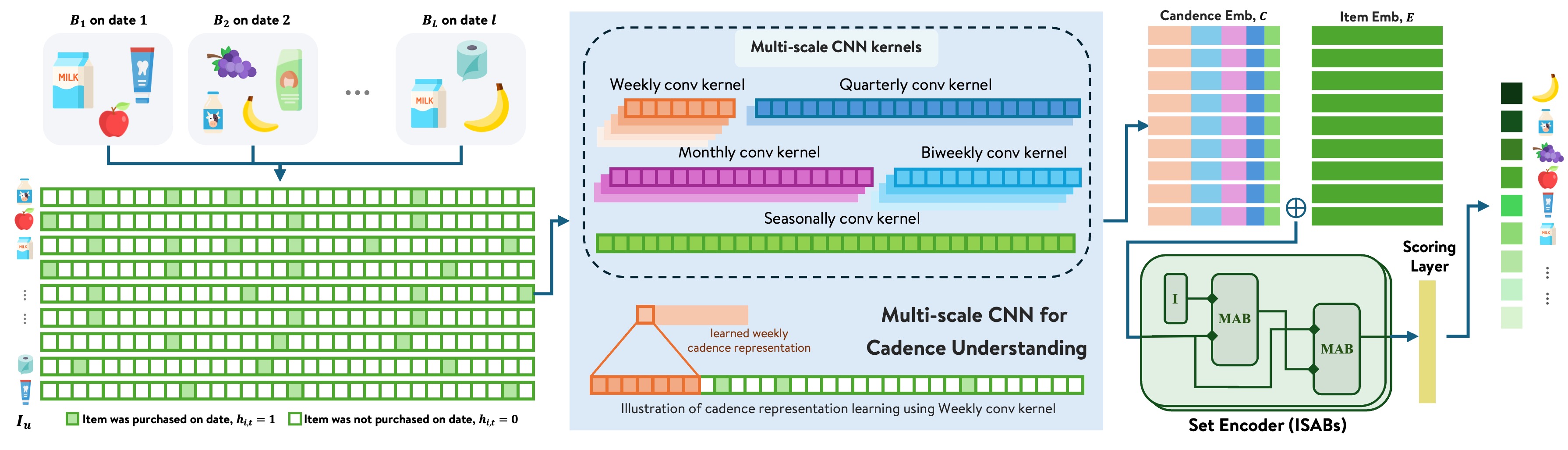}
    \caption{Overview of CASE. Item purchase histories are transformed into binary $T$-day calendar-time signals, enabling modeling of elapsed time and repurchase cadence. Shared multi-scale temporal convolutions learn population-wide recurring patterns and produce cadence embeddings $\mathbf{c}_i$, which are concatenated with item embeddings $\mathbf{e}_i$ and processed by Induced Set Attention Blocks (ISAB) to model cross-item dependencies. A final MLP outputs per-item repurchase scores.}    
    
    \label{fig:diagram}
\end{figure*}

In large-scale retail platforms with frequent replenishment behavior, a substantial fraction of items in a user’s next basket were previously purchased. Their repurchase timing often follows stable, item-specific cadences, such as milk purchased weekly and cleaning products purchased monthly. Thus, accurate timing is critical for user experience: recommending too early makes suggestions appear irrelevant, whereas recommending too late risks missing the purchase opportunity. This makes Next Basket Repurchase Recommendation (NBRR) a central task in production retail recommendation systems, where the goal is to predict which previously purchased items a user will need next.

Most neural NBRR methods model user history as an ordered sequence of basket events, where time is represented implicitly by basket index rather than by elapsed calendar time~\cite{yu2020predicting, sun2023generative, yu2023predicting, zhang2026diffnbr}. As a result, baskets on days 1, 8, and 36 are represented identically to baskets on days 1, 2, and 3 under the same three-step representation, and predictions are updated only when a new transaction occurs, leaving scores static between purchases and unable to reflect whether an item is overdue or not yet due. This creates a fundamental mismatch with production deployment: without explicit modeling of elapsed time, the model cannot meaningfully refresh its predictions between transactions and provides no adaptive signal for users whose purchase frequency changes over time. A related class of KNN-based methods models basket index with recency decay, retrieving similar users and aggregating their purchase patterns, and has shown strong performance on next basket repurchase recommendation benchmarks~\cite{hu2020modeling}. However, these approaches require computing user-to-user similarities at inference time, by comparing each query user against the entire user base, followed by aggregating signals from retrieved neighbors. At the scale of tens of millions of users, such per-query retrieval becomes computationally prohibitive for production deployment.

Some recent works have explored calendar-time representations for NBRR. One line represents item history as a binary time series and learns repurchase cycles via convolution~\cite{katz2024personalized}, showing the value of explicit cadence-aware modeling; however, it relies on user-specific convolutional filter parameterization, limiting scalability at production scale. A complementary approach~\cite{ranjan2025scalable} achieves scalable set-level modeling via permutation-equivariant aggregation over calendar-time membership, but does not explicitly extract multi-scale cadence patterns at the item level. 

In this paper, we propose \textbf{CASE} (\textbf{C}adence-\textbf{A}ware \textbf{S}et \textbf{E}ncoding for Large-Scale Next Basket Repurchase Recommendation), which explicitly models repurchase cadence in calendar time while remaining scalable at production scale. CASE applies shared multi-scale convolutional filters to capture item-level recurring patterns across weekly, biweekly, monthly, seasonal, and trend horizons. Cross-item dependencies within a user’s purchase history are encoded through induced set attention blocks (ISAB)~\cite{lee2019set}, reducing the quadratic complexity of full self-attention. Our contributions are:
\begin{itemize}[leftmargin=*,itemsep=0pt,topsep=2pt]
  \item We identify the basket-index formulation as a structural limitation of existing NBRR for production deployment, and motivate calendar-time cadence as a more suitable formulation.
  \item We propose CASE, combining shared multi-scale CNN-based cadence learning with induced set attention, requiring no per-user parameters and enabling efficient batch inference.
  \item We demonstrate up to 8.6\% relative Precision and 9.9\% Recall lift over a deployed production system on tens of millions of users and a large item catalog, along with consistent gains across three public benchmarks.
  \item Ablation shows cadence modeling dominates performance, with only modest degradation without item embeddings, reducing dependence on large, frequently refreshed embedding tables and enhancing production scalability.
\end{itemize}

\section{Model Architecture}
\label{sec:model}

% \textbf{Problem Formulation.}
Let $\mathcal{B}_u = (B_1, B_2, \ldots, B_L)$ denote the ordered basket sequence for user $u$, where $B_l \subseteq \mathcal{I}$ was purchased on calendar date $d_l$. The NBRR task ranks items in the repurchase history $\mathcal{I}_u = \bigcup_l B_l$ by their likelihood of appearing in $B_{L+1}$. Figure~\ref{fig:diagram} provides an overview of the CASE architecture.

\textbf{Calendar-Time History Representation.}
For each item $i \in \mathcal{I}_u$, we construct a binary purchase indicator $\mathbf{h}_i \in \{0,1\}^T$ over a rolling $T$-day calendar window, where $h_{i,t}{=}1$ if item $i$ was purchased on day $t$. This representation preserves the actual timing of purchases in calendar time, rather than compressing behavior into basket order alone, allowing the model to distinguish items with different repurchase cadences and support multi-scale learning of recurring replenishment patterns.

\textbf{Multi-Scale Temporal CNN.}
Shared multi-scale Conv1d filters are applied over $\mathbf{h}_i$ at five predefined kernel sizes: weekly ($w{=}7$), biweekly ($w{=}14$), monthly ($w{=}28$), seasonal ($w{=}91$), and trend ($w{=}182$), each with stride $w$ (non-overlapping windows). The $\lfloor T/w \rfloor$ activations per scale are concatenated across all scales and projected through two FC layers with ReLU to yield $\mathbf{c}_i \in \mathbb{R}^{d_c}$. This multi-resolution design captures both fine-grained repetition and slower-changing repurchase trends, allowing CASE to model temporal patterns across multiple horizons using population-wide shared weights.

\textbf{Induced Set Attention Encoding.} 
Since a user’s repurchase candidates naturally form an unordered set rather than a sequence, we adopt a Set Transformer encoder~\cite{lee2019set, cao2025s2srec2} to model cross-item dependencies without imposing an arbitrary item order. The combined representation $\mathbf{x}_i = [\mathbf{c}_i \,\|\, \mathbf{e}_i]$, where $\mathbf{e}_i \in \mathbb{R}^{d_e}$ is a learned item embedding, is then passed to the set encoder to model cross-item dependencies. We use Induced Set Attention Blocks (ISAB), which reduce $O(n^2)$ pairwise attention to $O(nK)$, where $n = |\mathcal{I}_u|$ is the number of candidate items for user $u$, via $K$ learnable induced points that first attend to the item set and then allow items to attend back. Two ISAB layers with $K{=}32$ and $H{=}4$ heads produce enriched representations $\mathbf{z}_i \in \mathbb{R}^{d_h}$.

\textbf{Scoring and Training.}
Each $\mathbf{z}_i$ is passed through a two-layer MLP to produce a scalar score $s_i$, and items are ranked by $s_i$ at inference. Training minimizes binary cross-entropy, treating items in $B_{l+1} \cap \mathcal{I}_u$ as positives and items in $\mathcal{I}_u \setminus B_{l+1}$ as negatives.

\begin{table*}[t]
\centering
\small
\caption{Comparisons on Top-$K$ performance for next basket repurchase recommendation. Higher is better. Best results per dataset and metric@K are in \textbf{bold}; second best are \underline{underlined}. We verify statistical significance using paired $t$-tests on metric scores ($\alpha = 0.05$), and consider differences significant when $p < 0.05$.}
\begin{tabular}{@{}ll|ccc|ccc|ccc|ccc@{}}
\toprule
\multirow{2}{*}{Datasets} & \multirow{2}{*}{Methods}
& \multicolumn{3}{c|}{$k{=}1$}
& \multicolumn{3}{c|}{$k{=}3$}
& \multicolumn{3}{c|}{$k{=}5$}
& \multicolumn{3}{c}{$k{=}10$} \\
\cmidrule(lr){3-5}
\cmidrule(lr){6-8}
\cmidrule(lr){9-11}
\cmidrule(lr){12-14}
& & Prec & Rec & NDCG
  & Prec & Rec & NDCG
  & Prec & Rec & NDCG
  & Prec & Rec & NDCG \\
\midrule

% ========================= TaFeng =========================
\multirow{6}{*}{TaFeng}
& PersonalTop
& 0.1771 & 0.0988 & 0.1771
& 0.1363 & 0.2100 & 0.2042
& 0.1104 & 0.2776 & 0.2263
& 0.0795 & 0.3839 & 0.2612 \\
& TIFUKNN
& 0.2146 & 0.1201 & 0.2146
& 0.1639 & 0.2646 & 0.2503
& 0.1332 & 0.3443 & 0.2769
& 0.0995 & 0.5007 & 0.3228 \\
& DNNTSP
& 0.2387 & 0.1632 & 0.2387
& \underline{0.1675} & \underline{0.3196} & 0.2869
& 0.1406 & \underline{0.4255} & 0.3275
& 0.1086 & 0.5809 & 0.3739 \\
& BERT4NBR
& 0.2316 & 0.1527 & 0.2316
& 0.1662 & 0.3035 & 0.2783
& 0.1399 & 0.3986 & 0.3120
& 0.1079 & 0.5753 & 0.3638 \\
& PIETSP
& \underline{0.2507} & \underline{0.1752} & \underline{0.2507}
& 0.1670 & 0.3165 & \underline{0.2903}
& \underline{0.1421} & 0.4223 & \underline{0.3317}
& \underline{0.1094} & \underline{0.5853} & \underline{0.3801} \\
& CASE
& \textbf{0.2897} & \textbf{0.1877} & \textbf{0.2897}
& \textbf{0.1944} & \textbf{0.3471} & \textbf{0.3260}
& \textbf{0.1539} & \textbf{0.4361} & \textbf{0.3559}
& \textbf{0.1157} & \textbf{0.5953} & \textbf{0.4021} \\
\midrule

% ========================= DC =========================
\multirow{6}{*}{DC}
& PersonalTop
& 0.3434 & 0.2879 & 0.3434
& 0.2233 & 0.5424 & 0.4489
& 0.1672 & 0.6559 & 0.4828
& 0.1059 & 0.7838 & 0.5000 \\
& TIFUKNN
& 0.3843 & 0.3245 & 0.3843
& 0.2498 & 0.6081 & \underline{0.5051}
& \underline{0.1843} & 0.7229 & 0.5387
& \underline{0.1140} & 0.8442 & \underline{0.5473} \\
& DNNTSP
& 0.3264 & 0.2685 & 0.3264
& 0.2269 & 0.5485 & 0.4441
& 0.1710 & 0.6696 & 0.4790
& 0.1085 & 0.8057 & 0.4931 \\
& BERT4NBR
& 0.3674 & 0.3102 & 0.3674
& 0.2392 & 0.5828 & 0.4820
& 0.1787 & 0.7009 & 0.5141
& 0.1121 & 0.8292 & 0.5183 \\
& PIETSP
& \underline{0.3886} & \underline{0.3291} & \underline{0.3886}
& \underline{0.2499} & \underline{0.6094} & 0.5023
& 0.1841 & \underline{0.7229} & \underline{0.5389}
& 0.1092 & \underline{0.8443} & 0.5467 \\
& CASE
& \textbf{0.3904} & \textbf{0.3296} & \textbf{0.3904}
& \textbf{0.2515} & \textbf{0.6113} & \textbf{0.5089}
& \textbf{0.1846} & \textbf{0.7230} & \textbf{0.5401}
& \textbf{0.1143} & \textbf{0.8468} & \textbf{0.5487} \\
\midrule

% ========================= Instacart =========================
\multirow{6}{*}{Instacart}
& PersonalTop
& 0.5019 & 0.1173 & 0.5019
& 0.4139 & 0.2593 & 0.4670
& 0.3574 & 0.3458 & 0.4556
& 0.2819 & 0.4821 & 0.4671 \\
& TIFUKNN
& \textbf{0.5489} & \textbf{0.1340} & \textbf{0.5489}
& \underline{0.4541} & \underline{0.2880} & \underline{0.5131}
& \underline{0.3930} & \underline{0.3878} & \underline{0.5037}
& \underline{0.3115} & \underline{0.5371} & \underline{0.5189} \\
& DNNTSP
& 0.4631 & 0.1025 & 0.4631
& 0.4002 & 0.2511 & 0.4447
& 0.3527 & 0.3480 & 0.4427
& 0.2842 & 0.4943 & 0.4616 \\
& BERT4NBR
& 0.3576 & 0.0812 & 0.3576
& 0.2875 & 0.1910 & 0.3276
& 0.2492 & 0.2608 & 0.3231
& 0.2047 & 0.3811 & 0.3397 \\
& PIETSP
& 0.5210 & 0.1222 & 0.5210
& 0.4424 & 0.2772 & 0.4951
& 0.3838 & 0.3773 & 0.4891
& 0.2986 & 0.5429 & 0.5155 \\
& CASE
& \underline{0.5478} & \underline{0.1325} & \underline{0.5478}
& \textbf{0.4551} & \textbf{0.2901} & \textbf{0.5135}
& \textbf{0.3989} & \textbf{0.3949} & \textbf{0.5086}
& \textbf{0.3155} & \textbf{0.5464} & \textbf{0.5241} \\
\midrule

% ========================= Proprietary =========================
\multirow{6}{*}{Proprietary}
& PersonalTop
& 0.3672 & 0.0921 & 0.3672
& 0.2744 & 0.1694 & 0.3251
& 0.2310 & 0.2205 & 0.3152
& 0.1766 & 0.2962 & 0.3121 \\
& TIFUKNN
& \underline{0.3856} & \underline{0.0971} & \underline{0.3856}
& \underline{0.3010} & \underline{0.1913} & \underline{0.3534}
& \underline{0.2615} & \underline{0.2550} & \underline{0.3503}
& \textbf{0.2038} & \textbf{0.3547} & \textbf{0.3543} \\
& DNNTSP
& 0.2661 & 0.0614 & 0.2661
& 0.2154 & 0.1269 & 0.2479
& 0.1852 & 0.1703 & 0.2420
& 0.1474 & 0.2510 & 0.2456 \\
& BERT4NBR
& 0.2288 & 0.0674 & 0.2288
& 0.1522 & 0.1132 & 0.1925
& 0.1275 & 0.1453 & 0.1878
& 0.0979 & 0.2028 & 0.1908 \\
& PIETSP
& 0.3803 & 0.0959 & 0.3803
& 0.2913 & 0.1814 & 0.3431
& 0.2422 & 0.2311 & 0.3301
& 0.1822 & 0.3094 & 0.3245 \\
& CASE
& \textbf{0.3871} & \textbf{0.1007} & \textbf{0.3871}
& \textbf{0.3046} & \textbf{0.1921} & \textbf{0.3571}
& \textbf{0.2661} & \textbf{0.2550} & \textbf{0.3509}
& \underline{0.2017} & \underline{0.3448} & \underline{0.3514} \\
\bottomrule
\end{tabular}
\label{tab:main_results}
\end{table*}

\section{Experimental Setting}

\subsection{Datasets}

We evaluate on four datasets in the grocery domain, chosen to cover a range of catalog sizes, basket densities, and user scales.

\textbf{Instacart}\footnote{\url{https://www.kaggle.com/datasets/yasserh/instacart-online-grocery-basket-analysis-dataset}} is a public online grocery dataset with 18,739 users, 37,522 products, and an average of 16.7 baskets per user with 10.07 items each. \textbf{TaFeng}\footnote{\url{https://www.kaggle.com/datasets/chiranjivdas09/ta-feng-grocery-dataset}} is a Chinese grocery store transaction dataset with 7,227 users and 18,703 items, whose shorter histories (7.52 baskets per user) and smaller baskets (6.58 items) make it a challenging setting for cadence models. \textbf{DC} is derived from the Dunnhumby ``Carbo-Loading'' database\footnote{\url{https://www.dunnhumby.com/source-files/}} and contains 123,935 users but only 852 distinct products, with very small baskets (1.60 products per basket). \textbf{Proprietary} data is randomly sampled from our internal large-scale grocery dataset, having 10,308 users across 88,812 items, with rich histories of 11.72 items per basket and 43 baskets per user on average. 

For all datasets, we perform a user-level train–test split and adopt a leave-one-out evaluation where each user’s last basket serves as the target basket. Candidates are restricted to items previously purchased by the user. We preserve calendar timestamps when available (TaFeng and Proprietary); otherwise (Instacart and DC), we reconstruct relative dates from inter-order gaps (e.g., days-since-prior-order) to build the $T$-day binary history representation.

\subsection{Baselines}

We compare CASE against five baselines: \textbf{PersonalTop}, a simple personalized frequency baseline that ranks each user’s previously purchased items by historical purchase count; \textbf{TIFUKNN}~\cite{hu2020modeling}, a KNN-based method that builds temporally decayed item vectors for each user and aggregates scores from similar users, achieving strong repurchase performance but with inference-time user retrieval that is not scalable in production; \textbf{DNNTSP}~\cite{yu2020predicting}, which applies GNN aggregation over an item co-occurrence graph together with basket-index-based temporal attention; \textbf{BERT4NBR}~\cite{li2023masked}, an adaptation of BERT4Rec to the Next Basket Recommendation (NBR) setting, using bidirectional self-attention over basket-indexed purchase sequences; and \textbf{PIETSP}~\cite{ranjan2025scalable}, which performs scalable set-level aggregation over time-step-based item histories using permutation-equivariant mean pooling, but without explicit multi-scale temporal feature extraction at the item level.

\begin{table*}[t]
\centering
\small
\caption{Ablation study on the Instacart dataset. Higher is better. Best results are in \textbf{bold}; second best are \underline{underlined}.}
\label{tab:ablation}
\begin{tabular}{l|ccc|ccc|ccc|ccc}
\toprule
\multirow{2}{*}{Model Components}
& \multicolumn{3}{c}{$k{=}1$}
& \multicolumn{3}{|c}{$k{=}3$}
& \multicolumn{3}{|c}{$k{=}5$}
& \multicolumn{3}{|c}{$k{=}10$} \\
\cmidrule(lr){2-4}
\cmidrule(lr){5-7}
\cmidrule(lr){8-10}
\cmidrule(lr){11-13}
& Prec & Rec & NDCG
& Prec & Rec & NDCG
& Prec & Rec & NDCG
& Prec & Rec & NDCG \\
\midrule

CASE w/o CNN
& 0.3333 & 0.0740 & 0.3333
& 0.2666 & 0.1793 & 0.3046
& 0.2382 & 0.2519 & 0.3065
& 0.2023 & 0.3784 & 0.3297 \\

CASE w/o Set Encoder
& 0.5232 & 0.1263 & 0.5232
& 0.4485 & 0.2897 & 0.5038
& 0.3947 & 0.3915 & 0.5000
& 0.3122 & 0.5396 & 0.5152 \\

CASE w/o Item Embedding
& 0.5390 & 0.1281 & 0.5390
& 0.4488 & 0.2848 & 0.5057
& 0.3919 & 0.3847 & 0.4985
& 0.3124 & 0.5360 & 0.5161 \\

CASE w/ PermEqMean
& \underline{0.5414} & \underline{0.1326} & \underline{0.5414}
& \underline{0.4538} & \underline{0.2901} & \underline{0.5115}
& \underline{0.3983} & \underline{0.3917} & \underline{0.5062}
& \underline{0.3147} & \underline{0.5463} & \underline{0.5225} \\

CASE (w/ ISAB)
& \textbf{0.5478} & \textbf{0.1325} & \textbf{0.5478}
& \textbf{0.4551} & \textbf{0.2901} & \textbf{0.5135}
& \textbf{0.3989} & \textbf{0.3949} & \textbf{0.5086}
& \textbf{0.3155} & \textbf{0.5464} & \textbf{0.5241} \\

\bottomrule
\end{tabular}
\end{table*}

\subsection{Evaluation and Implementation}

We report three metrics at $k \in \{1, 3, 5, 10\}$. \textbf{Precision@$k$} measures the fraction of recommended items that are relevant, which is the primary metric in user-facing recommendation systems; \textbf{Recall@$k$} measures the fraction of all relevant items captured in the top-$k$ list; \textbf{NDCG@$k$} is a ranking-aware metric that assigns higher scores to positive items appearing higher in the list. 

CASE is implemented in PyTorch with item embedding dimension $d_e{=}128$, CNN output dimension $d_c{=}128$, ISAB hidden dimension $d_h{=}256$, $K{=}32$ induced points, and $H{=}4$ attention heads. We train for 30 epochs using the Adam optimizer with learning rate $10^{-3}$, weight decay $10^{-5}$, and batch size 64. Dropout of 0.1 is applied after each ISAB layer and in the MLP scorer. For DNNTSP, BERT4NBR, and PIETSP, we optimized the models using the same optimization setup as CASE, running for 30 epochs and selecting the best model based on validation performance.

\section{Experimental Results}

Table~\ref{tab:main_results} compares CASE against five baselines across four datasets. CASE achieves the best or near-best performance across all datasets and metrics. The primary offline competitor is TIFUKNN, a well-established strong baseline for repurchase recommendation. A consistent pattern across datasets is that CASE shows the largest gains in sparse settings. On TaFeng (sparse histories) and DC (small baskets), CASE leads, with PIETSP ranking second, confirming that calendar-time cadence provides a stronger signal when transaction frequency is low and basket context is limited. In such settings, neighbor-based aggregation becomes less reliable, whereas CASE’s shared temporal encoding continues to generalize across users. On richer datasets (Instacart and Proprietary), CASE remains competitive with TIFUKNN and slightly outperforms it on most cutoffs, which is notable given TIFUKNN’s effectiveness as a repurchase baseline~\cite{li2023next}. These results are consistent with our ablation findings, which show that cadence modeling is the dominant contributor to performance. DNNTSP and BERT4NBR consistently lag across DC, Instacart, and Proprietary, sometimes even below PersonalTop, a simple frequency-based baseline. This suggests that basket-index-based representations impose a structural limitation: more complex graph or transformer architectures cannot compensate for the lack of explicit calendar-time modeling needed to capture true repurchase timing dynamics.

Besides recommendation quality, CASE also keeps parameterization independent of the user population. For a user with $n$ repurchase candidates and a $T$-day horizon, multi-scale temporal encoding and induced set attention incur $O(n(T+K))$ complexity, where $K$ is the number of learnable induced points in ISAB. TIFUKNN, by contrast, requires $O(|\mathcal{U}| \cdot d)$ per-query computation to retrieve and aggregate neighbor histories (where $|\mathcal{U}|$ is the number of users and $d$ is the embedding dimension), which is infeasible at the scale of tens of millions of users. This gap is also reflected empirically: on Instacart test data with 3,744 users, CASE achieves substantially lower inference time than TIFUKNN (19s vs.\ 318s), while also delivering stronger recommendation quality.

Overall, CASE is the only approach in our comparison that is both competitive with the strongest offline baseline and practical for deployment at production scale.

\subsection{Ablation Study}

The ablation study is conducted on the Instacart dataset and isolates the contribution of each architectural component, shown in Table~\ref{tab:ablation}.

\begin{figure}[h]
    \centering    
    \includegraphics[width=0.8\linewidth]{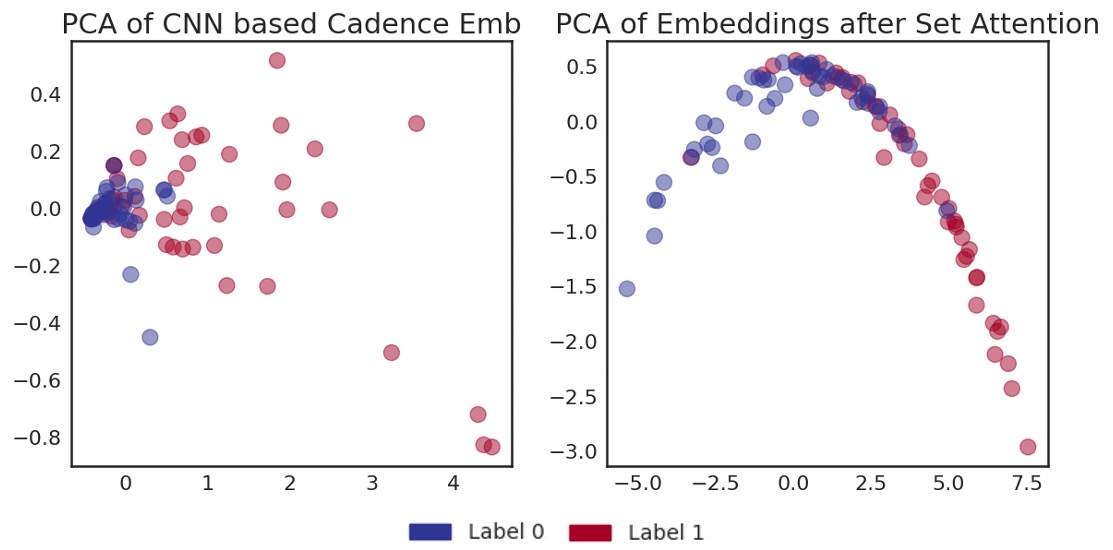}
    \caption{PCA visualization of item embeddings (50 positive/negative samples each). Left: cadence embeddings $\mathbf{c}_i$; Right: set-encoded representations $\mathbf{z}_i$. CNN induces cadence-based separation, while ISAB refines the boundary through cross-item interaction.}
    \label{fig:embedding}
\end{figure}

\textbf{Temporal CNN is the dominant component.} 
Removing the multi-scale CNN leads to the largest performance drop across all metrics. This confirms that calendar-time cadence encoding provides the primary discriminative signal in CASE. Figure~\ref{fig:embedding} further supports this observation: the temporal CNN separates positives and negatives by cadence phase, while ISAB preserves this structure and sharpens the boundary through cross-item interaction.

\textbf{Set Attention provides consistent gains across $k$.}
Removing the set encoder reduces performance at every cutoff. This suggests that ISAB captures cross-item co-purchase dependencies that complement temporal cadence signals throughout the ranked list.

We further compare ISAB to a permutation-equivariant mean pooling encoder (PermEqMean)~\cite{zaheer2017deep} to assess whether attention-based interaction provides benefits beyond simple set aggregation. PermEqMean achieves competitive performance, indicating that set-level encoding is effective. ISAB further yields consistent gains, suggesting that induced-point attention captures richer cross-item dependencies than simple permutation-invariant aggregation.

\textbf{Item Embedding plays a complementary role.}
Removing item embeddings degrades performance modestly, indicating that semantic identity provides additional signal beyond cadence. The small gap confirms that calendar-time encoding is the primary driver of performance. This is significant in production: it reduces reliance on large, frequently refreshed embedding tables, improving scalability as the catalog grows.

\section{Production Experimental Results}

We compare CASE against our production learning-to-rank model with engineered repurchase features, using the same data scale and candidate pipeline. We report results at $k{\in}\{5, 10, 20\}$, which reflect the slate sizes shown to users. As shown in Table~\ref{tab:lift}, CASE achieves consistent relative lifts of 6.8--8.6\% in Precision and 7.9--9.9\% in Recall, along with corresponding improvements in NDCG.

\begin{table}[h]
\centering
\small
\caption{Relative lift (\%) of CASE over production model.}
\begin{tabular}{c|ccc}
\toprule
k & Precision & Recall & NDCG \\
\midrule
5  & +8.63\%  & +9.90\%  & +10.46\% \\
10 & +6.78\%  & +7.95\%  & +9.40\% \\
20 & +5.27\%  & +6.32\%  & +8.75\% \\
\bottomrule
\end{tabular}
\label{tab:lift}
\end{table}

\section{Conclusion}

CASE uses shared multi-scale temporal filters and induced set attention, ensuring that inference scales linearly with candidate size and remains independent of the total user population. This design allows CASE to combine strong offline recommendation quality with practical deployability at production scale, without introducing per-user parameterization or additional serving complexity. In particular, the model satisfies key production constraints on scalability and infrastructure while preserving efficient inference over large candidate sets. An online A/B test is planned to further validate whether these offline gains translate into improvements on user-facing metrics.

While CASE demonstrates strong performance across benchmarks and production-scale evaluations, several directions remain for future study. First, CASE assumes relatively stable historical cadences. Its ability to handle disrupted or non-stationary signals, such as out-of-channel purchases or changing user routines, needs further investigation. Second, the current design uses fixed multi-scale kernels and a fixed temporal horizon. Future work could explore more adaptive temporal parameterizations and assess sensitivity to architectural choices. Finally, although CASE performs strongly on sparse datasets such as TaFeng and DC, it would be valuable to further understand its behavior in more extreme boundary regimes, including items with very long repurchase cycles, newly introduced items, and users with extremely limited purchase histories.

\section*{Main Presenter Bio}

\noindent \textbf{Yanan Cao} 
is a Senior Data Scientist at Walmart Global Tech, working on large-scale retail recommendation systems. Her work focuses on next basket recommendation, purchase interval prediction, and temporal modeling for production-scale recommender systems using transformer-based deep learning and large language model reasoning. Her interests lie in scalable learning systems, and translating research innovations into real-world deployment.

\noindent \textbf{Ashish Ranjan} 
is a Staff Data Scientist at Walmart Global Tech, leading large-scale personalization and ranking systems. His work spans heterogeneous-set modeling, unified ranking across diverse content, and integration of large language models into production e-commerce workflows. He has led end-to-end deployments, from modeling to real-time inference infrastructure, serving hundreds of millions of users. His interests include scalable IR systems and the deployment of learning architectures at industrial scale.

\balance

% \clearpage
\bibliographystyle{ACM-Reference-Format}
\bibliography{bibfile}

\end{document}